\documentclass[preprint2]{aastex62}
\newcommand{\roma}[1]{\uppercase\expandafter{\romannumeral#1}}
\newcommand{\speed}[1]{#1 km~s${}^{-1}$}
\newcommand{\accel}[1]{#1 km~s${}^{-2}$}

\newcommand{\nfig}[1]{Figure~\ref{#1}}

\usepackage{amsmath}
\usepackage{subfigure}
\usepackage{hyperref}

\usepackage{lineno}
\usepackage{graphicx}
\usepackage{natbib}
\usepackage{amssymb,txfonts}
\usepackage{multirow}
\usepackage{array}
\usepackage{sidecap}
\usepackage{hyperref}
\usepackage{epstopdf}
\usepackage{footnote}
\usepackage{tabularx}
\usepackage{booktabs}
\usepackage{CJKnumb}
\usepackage{bm}
\usepackage{appendix}
\shortauthors{Yao et al.}

\begin{document}
\title{Formation and Eruption of Hot Channel Magnetic Flux Rope in Nested Double Null Magnetic System}
\correspondingauthor{Yuandeng Shen}
\email{ydshen@hit.edu.cn}
\author{Surui Yao}
\affiliation{Yunnan Observatories, Chinese Academy of Sciences,  Kunming, 650216, China}
\affiliation{University of Chinese Academy of Sciences, Beijing 100049, China}
\author[0000-0001-9493-4418]{Yuandeng Shen}
\affiliation{Shenzhen Key Laboratory of Numerical Prediction for Space Storm, School of Aerospace, Harbin Institute of Technology, Shenzhen 518055, China}
\author{Chengrui Zhou}
\affiliation{Yunnan Observatories, Chinese Academy of Sciences,  Kunming, 650216, China}
\affiliation{University of Chinese Academy of Sciences, Beijing 100049, China}
\author{Dongxu Liu}
\affiliation{Yunnan Observatories, Chinese Academy of Sciences,  Kunming, 650216, China}
\affiliation{University of Chinese Academy of Sciences, Beijing 100049, China}
\author[0000-0001-9374-4380]{Xinping Zhou}
\affiliation{College of Physics and Electronic Engineering, Sichuan Normal University, Chengdu 610068, China}

\begin{abstract}
The coronal magnetic topology significantly affects the outcome of magnetic flux rope (MFR) eruptions. The recently reported nested double null magnetic system remains unclear as to how it affects MFR eruptions. Using observations from the New Vacuum Solar Telescope and the Solar Dynamics Observatory, we studied the formation and successful eruption of a hot channel MFR from NOAA active region AR12173 on 2014 September 28. We observed that a hot channel MFR formed and erupted as a coronal mass ejection (CME), and the magnetic field of the source region was a nested double null magnetic system in which an inner magnetic null point system was nested by an outer fan-spine magnetic system. Observational analysis suggests that the origin of the MFR was due to magnetic reconnection at the inner null point, which was triggered by the photospheric swirling motions. The long-term shearing motion in the source region throughout around 26 hours might accumulate enough energy to power the eruption. Since previous studies showed that MFR eruptions from nested double null magnetic systems often result in weak jets and stalled or failed eruptions, it is hard to understand the generation of the large-scale CME in our case. A detailed comparison with previous studies reveals that the birth location of the MFR relative to the inner null point might be the critical physical factor for determining whether an MFR can erupt successfully or not in such a particular nested double null magnetic system.
	
\end{abstract}

\section{Introduction}
A magnetic flux rope (MFR) is a group of helical magnetic field lines wrapping around a common axis, which hints that magnetic field lines inside an MFR have the same central axis and are rooted around similar locations \citep[e.g.,][]{2020RAA....20..165L}. Observationally, the MFRs can contain filaments \citep[e.g.,][]{1998A&A...335..309A, 2010SSRv..151..333M}, composed of relatively cold and dense plasma, which are typically seen as dark and elongated structures in absorption in H$\alpha$ and extreme ultraviolet (EUV) wavelengths on the solar disk, while appearing as bright features off the solar disk. On the solar disk, MFRs may appear as sigmoids with a forward or reversed sigmoidal shape, observed in EUV and soft X-ray (SXR) wavelengths \citep[e.g.,][]{2009ApJ...703.1766S, 2010ApJ...708..314A}. Also, the hot channels are widely accepted to be MFRs \citep[e.g.,][]{2012NatCo...3..747Z, 2013ApJ...763...43C, 2014ApJ...789...93C}. Hot channels are visible only at AIA high-temperature passbands (e.g., AIA 94 and 131 \AA), while invisible in the other low-temperature passbands. An MFR is an ideal carrier of magnetic free energy, whose eruption is typically in association with violent solar activities, such as flares \citep[e.g.,][]{2015NatCo...6.7008W}, large-scale magnetohydrodynamic (MHD) waves \citep[e.g.,][]{2018ApJ...853....1S, 2019ApJ...873...22S, 2022SoPh..297...20S, 2021SoPh..296..169Z}, CMEs \citep[e.g.,][]{2011ApJ...732L..25C, 2011RAA....11..594S, 2012ApJ...745..164S} and solar storms \citep[e.g.,][]{2018SpWea..16..216M}.

The initiation and eruption mechanisms of MFRs have attracted much attention. To date, two types of physical mechanisms are frequently used to explain the initiation and eruption of MFRs. One is the ideal MHD instabilities without dissipation process, including the kink instability \citep[e.g.,][]{2004A&A...413L..27T, 2003ApJ...595L.135J} and the torus instability \citep[e.g.,][]{2006PhRvL..96y5002K}. For kink instability, it will occur if the magnetic twist of the MFR exceeds a critical value \citep[e.g.,][]{2004A&A...413L..27T}. The torus instability is determined by the competition mainly between the Lorentz-self force (hoop force; pointed outwards) and the strapping force (generated by the external poloidal field and pointed inwards). With the radius of the current ring expanding rapidly, if the inward strapping force decreases faster than the outward hoop force, the system will become unstable, which refers to the torus instability \citep[e.g.,][]{2006PhRvL..96y5002K}. Another type is reconnection-based models. For example, in the so-called tether-cutting model \citep[e.g.,][]{2001ApJ...552..833M, 2014ApJ...797L..15C, 2024ApJ...964..125S}, two sets of magnetic field lines can be formed above the polarity inversion line (PIL) of a bipolar region due to the magnetic reconnection in the current sheet between the two groups of J-shaped sheared magnetic field lines. While the outward-moving magnetic field lines form a long MFR, the downward-moving reconnected field lines form the hot flare arcade. Another well-known reconnection-based model under a bipolar field is called the standard flare model \citep[e.g.,][]{1964NASSP..50..451C, 1966Natur.211..695S, 1974SoPh...34..323H, 1976SoPh...50...85K}. In standard flare models, when a preexisting MFR erupts upwards, magnetic field inflow beneath it creates a current sheet where reconnection rapidly converts magnetic energy into plasma heating and particle acceleration. \cite{2000JGR...105.2375L} proposed a two-dimensional hybrid CME model including an MFR, in which the system is initiated by the loss of equilibrium of the pre-eruption configuration, and stretching of the background strapping magnetic field lines can form a current sheet underneath the rising MFR. The fast magnetic reconnection within the current sheet can accelerate the eruption of the MFR to form a CME, leaving behind two conjugated flare ribbons connected by a group of hot flare arcades in the eruption source region.

Compared with a bipolar magnetic field, \cite{1999ApJ...510..485A} proposed the well-known magnetic breakout model to emphasize the importance of multipolar magnetic fields in forming CMEs and flares. In the magnetic breakout model, a magnetic null point exists between the core and the overlying fields. The magnetic null point where the magnetic field vanishes is a crucial element in the magnetic breakout model, and it can be the location for the formation of a current sheet \citep[e.g.,][]{2014masu.book.....P}. Reconnection at the null point can remove the enveloping field above the core field, allowing the core field to further expand \citep[e.g.,][]{1999ApJ...510..485A, 2004ApJ...614.1028M, 2012ApJ...760...81K}. Usually, the magnetic null point is associated with the discontinuity in the magnetic field; it often exists in regions where magnetic field lines show strong gradients that can be identified as quasi-separatrix layers \citep[QSLs;][]{1995JGR...10023443P}. Generally, regions with high-degree $Q$-values often denote the possible reconnection locations \citep[e.g.,][]{2012A&A...541A..78P}, and intense current will be generated along QSLs due to the large gradients of the magnetic field \citep[e.g.,][]{2014masu.book.....P}. Calculating the squashing factor $Q$ in a magnetic field is often used to find locations where magnetic reconnection might occur. In principle, regions where the squashing factor $Q$ is much greater than 2 are usually regarded as QSLs \citep{2002JGRA..107.1164T}. 

Simulation and observation evidence suggest that the magnetic breakout model is a universal mechanism for explaining solar eruptions across multiple scales, from large-scale CMEs to small-scale solar jets \citep[e.g.,][]{2017Natur.544..452W, 2018ApJ...852...98W, 2021ApJ...909...54W, 2018ApJ...854..155K, 2021ApJ...907...41K, 2019ApJ...886L..34L, 2021ApJ...923...45Z}. According to this model, through shearing the field lines along the PIL beneath the null point, the breakout system is energized and a twisted flux rope might be created underneath the central strapping field prior to the onset of the breakout phase \citep[e.g.,][]{1999ApJ...510..485A, 2017Natur.544..452W, 2018ApJ...852...98W}. During the slow rise of the flux rope, the null point becomes increasingly compressed as the core confining field lines of the flux rope expand upwards towards it, and then a breakout current sheet (BCS) forms, and magnetic reconnection within it will remove some strapping field lines above the rising flux rope. Hence, as the flux rope rises, a flare current sheet (FCS) forms below it, and the flare reconnection takes place and accelerates the breakout reconnection in a positive feedback process. For breakout jets, the corresponding configuration is often called a fan-spine magnetic system, which is composed of the spine lines inside/outside the dome-shaped fan with a coronal null point among them \citep[e.g.,][]{2013ApJ...778..139S, 2017Natur.544..452W, 2018ApJ...852...98W, 2019ApJ...885L..11S, 2019ApJ...871....4H, 2019ApJ...874..146H, 2021RSPSA.47700217S}.

Remarkably, eruptions in a nested magnetic flux system containing two null points were observed and analyzed in previous studies \citep[e.g.,][]{2019ApJ...871....4H, 2019ApJ...886L..34L, 2024ApJ...966...27K}, which can be named as a nested double null magnetic system. Especially, \cite{2024ApJ...966...27K} reported a detailed investigation of an MFR eruption in a nested double null magnetic system, and the MFR underwent a stalled eruption within a nested fan-spine system. In their case, an MFR formed and accelerated quickly in a nested double null magnetic system but stopped and stayed within the large-scale fan-spine magnetic system, which suggests that the nested double null magnetic system adds the complexity of the eruption process.

Similar to \cite{2024ApJ...966...27K}, our study also focused on the eruption in a nested double null magnetic system. The main difference between to \cite{2024ApJ...966...27K} is that in our case, the MFR escaped successfully and caused a CME. Thus, a comparison study relative to \cite{2024ApJ...966...27K} is necessary to investigate the physical reasons for determining whether a CME can occur in such a nested double null magnetic system, and this can also help deepen our understanding of the eruptions in the nested double null magnetic system and the magnetic breakout mechanism. The present study mainly focuses on how the CME formed from a complicated nested double null magnetic system and aims to answer the question of why the CME can occur in our case but not in \cite{2024ApJ...966...27K}. Instruments and data details are introduced in Section \ref{sec:2}. Results are presented in Section \ref{sec:3}. Discussions and a summary are given in Section \ref{sec:4}.

\begin{figure*}
\epsscale{0.85}
\figurenum{1}
\plotone{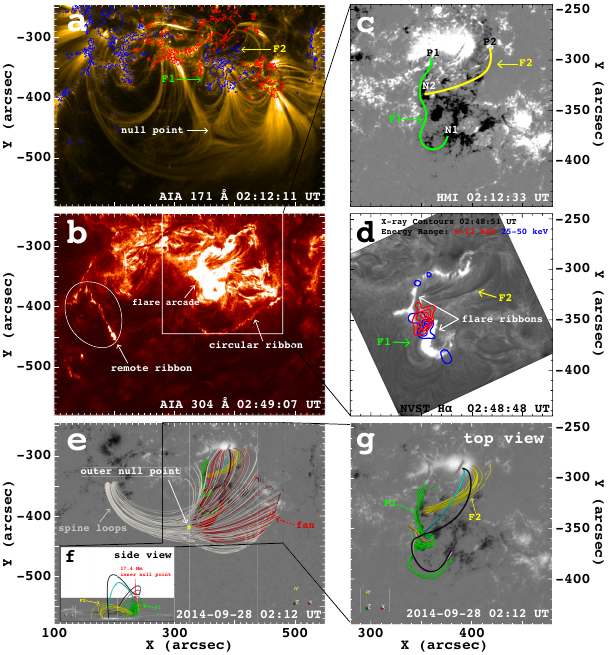}
\caption{Overview of this event on 2014 September 28. Panels (a) is AIA 171 \AA\ image overlaid by HMI magnetogram contours ($\pm$ 300 G, red = positive, blue = negative). Two filaments (F1 and F2; indicated by the green and yellow arrows, respectively) exist under a large-scale fan-spine magnetic flux system within a null point, noted by a white arrow. Panel (b) is AIA 304 \AA\ image and displays the eruption signatures of the source region. The white rectangle in panel (b) indicates the field of view (FOV) of panels (c) and (d). Panel (c) presents the locations of the two filaments (F1 and F2; indicated by the green and yellow lines) on HMI magnetogram. In panel (d), H$\alpha$ image acquired by the NVST shows the filaments were stable during the entire process, which is overlaid with the {\em RHESSI} images for 6 -- 12 (red) and 25 -- 50 (blue) keV for 20$\%$, 45$\%$, 75$\%$ and 95$\%$ of the maximum emission. The white arrows in panel (d) indicate the two flare ribbons. Panel (e) is the reconstructed 3D magnetic topology of the source region revealed by NLFFF extrapolation at 02:12 UT on 2014 September 28. The gray and red curves indicate the outer fan-spine magnetic system. The black box in panel (e) indicates the FOV of panels (f) and (g) and has the same range of the white box in panel (b). Panel (f) shows the side view of the inner magnetic null point system (indicated by the black, pink, cyan and orange curves), containing an inner null point (17.4 Mm above the photosphere). Panel (g) is the top view of the inner magnetic null point system. The green and yellow curves in extrapolation results indicate the two observed filaments.
\label{fig1}}
\end{figure*}

\section{Instruments and Data}
\label{sec:2} 
We used the extreme ultraviolet (EUV) data taken by the Atmospheric Imaging Assembly \cite[AIA;][]{2012SoPh..275...17L} and the magentograms taken by the Helioseismic and Magnetic Imager \cite[HMI;][]{2012SoPh..275..229S} onboard the {\em Solar Dynamics Observatory} \cite[{\em SDO};][]{2012SoPh..275....3P}. This event was also observed by the X-ray Telescope \citep[XRT;][]{2007SoPh..243...63G} onboard the {\em Hinode} \citep{2007SoPh..243....3K}. High temporal and spatial resolution H$\alpha$ observations taken by the New Vacuum Solar Telescope \cite[NVST;][]{2014RAA....14..705L}, soft X-ray (SXR) and hard X-ray (HXR) data recorded respectively by the {\em Geostationary Operational Environmental Satellites} ({\em GOES}) and the {\em Reuven Ramaty High Energy Solar Spectroscopic Imager} \citep[{\em RHESSI};][]{2002SoPh..210....3L}. White-light coronal observations taken by the Large Angle and Spectrometric Coronagraph \cite[LASCO;][]{1995SoPh..162..357B} onboard the {\em Solar and Heliospheric Observatory} ({\em SOHO}) are also used.

The AIA image has a temporal cadence of 12 s and a spatial resolution of 1\arcsec.2, which provides quasi-simultaneous multi-wavelength ultraviolet and EUV images to probe the solar atmosphere from the photosphere to the corona. The HMI provides data, including the photospheric Doppler speed, line-of-sight (LOS) magnetograms, and vector magnetograms with a spatial resolution of 1\arcsec. The cadences of the Doppler speed and LOS magnetograms are both 45 s, while that of the vector magnetograms is 720 s. The NVST took the H$\alpha$ center images with a cadence of 12 s and a spatial resolution of 0\arcsec.33. In addition, the LASCO provides the white-light images of the outer corona from 2 to 32 R$_\odot$.

\section{Results}
\label{sec:3}  
On 2014 September 28, we observed a nested double null magnetic system in NOAA AR12173, where a large-scale fan-spine magnetic system hosts a small-scale magnetic null point system. A hot channel erupted from below the large-scale fan-spine magnetic system, resulting in a {\em GOES} M5.1 flare whose start and peak times were at about 02:39 UT and 02:58 UT, respectively. In addition, LASCO observed a slow CME traveling toward the southwest at an average speed of about \speed{215}.

\begin{figure*}
\epsscale{0.85}
\figurenum{2}
\plotone{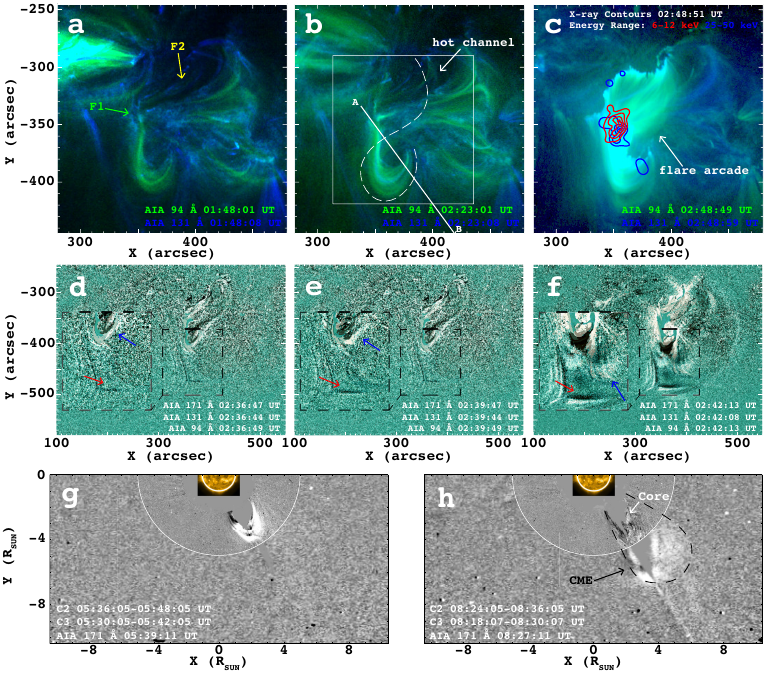}
\caption{Evolution of the source region on 2014 September 28. Panels (a) -- (c) are composite images of AIA 94 \AA\ (green) and 131 \AA\ (blue) during different evolution phases: the pre-eruption, eruption, and post-eruption phases. In Panel (b), the dashed white curve indicates the axis of the hot channel MFR, which is shifted to the left to avoid covering the MFR due to its weak radiation. The white line in panel (b) denotes the path of the time-distance diagram in \nfig{fig3}(a), from A to B. The white box in panel(b) indicates the integration region of the normalized light curve of AIA 131 \AA\ in \nfig{fig3}(b). Panels (d) -- (f) are differential composite images of AIA 94 \AA, 131 \AA\ and 171 \AA. The blue and red arrows mark the hot channel and the outer null point of the large-scale fan-spine magnetic system, respectively. Panels (g) and (h) show the running difference images of a CME observed by LASCO/C2 and C3, with AIA 171 \AA\ images. A dashed black curve marks the bright front of the CME, and the bright core is denoted by the white arrow in (h). The animation covers a duration of about 90 minutes from 02:00 UT to 03:30 UT on 2014 September 28.
\\
(An animation of this figure is available)
\label{fig2}}
\end{figure*}

\subsection{Overview}
\nfig{fig1} shows the overview of the event on 2014 September 28. As displayed in \nfig{fig1}(a), a large-scale fan-spine magnetic system can be observed directly in the AIA 171 \AA\ image at 02:12:11 UT before the eruption. During the violent eruption of the event, we observed a circular flare ribbon and a bright remote ribbon as indicated by the white arrows in \nfig{fig1}(b). Under the large-scale fan-spine magnetic system, there were two filaments (F1 and F2) as indicated by the green and yellow arrows in \nfig{fig1}(a) and (d). The projection positions of the two filaments are overlaid on the HMI line-of-sight (LOS) magnetogram as indicated by green and yellow curves in \nfig{fig1}(c). It is clear that the two ends of F1 were rooted in the two main magnetic regions of opposite polarities (P1 and N1; see the green curve in \nfig{fig1}(c)), while those of F2 were rooted in two small satellite magnetic regions at the periphery of the two main magnetic poles (P2 and N2; see the yellow curve in \nfig{fig1}(c)). Since the AR12173 was located in the southwest of the solar disk, F1 and F2 may deviate from the photospheric polarity inversion line (PIL) due to the projection effect. F1 and F2 can be observed clearly in the NVST H$\alpha$ image (see \nfig{fig1}(d)). In addition, the contours of {\em RHESSI} sources at energy bands of 6 -- 12 (red) and 25 -- 50 (blue) keV were also overlaid on the H$\alpha$ image. One can see that the hard X-ray {\em RHESSI} sources at the energy band of 25 -- 50 keV reveal not only the loop-top source (the middle one) but also the footpoint sources (the upper and bottom ones located on the flare ribbons). The two filaments remained stable at their original site throughout the whole eruption process (denoted by the green and yellow arrows in \nfig{fig1}(d)). This raises the question of what caused the observed flare and CME. Therefore, we alternatively investigated the pre-eruption magnetic configuration below the large-scale fan-spine magnetic system to determine the eruption mechanism and interpret the associated flare and CME.

We extrapolated the three-dimensional coronal magnetic field of the eruption source region using the nonlinear force-free magnetic field (NLFFF) optimization method based on the photospheric vector magnetic field as input \citep{2000ApJ...540.1150W}, to study the pre-eruption magnetic configuration of the source region. The coronal magnetic field satisfying equations, i.e., \bm{$\bigtriangledown$} $\times$ \bm{$B$} $= \alpha$\bm{$B$} and \bm{$\bigtriangledown$} $\cdot$ \bm{$B$} $= 0$ meets the force-free criteria, where $\alpha = \alpha(x)$ is a function of position along each magnetic field line. When $\alpha$ varies in space, the extrapolated field becomes nonlinear. We used the vector magnetic field observed by the HMI at 02:12 UT before the onset of the event as the bottom boundary condition to extrapolate the three-dimensional coronal magnetic field. We plotted some key magnetic field lines in \nfig{fig1}(e) -- (g) to show the basic magnetic connectivity and topology of the eruption source region. Interestingly, the extrapolated results show an inner magnetic null point system situated within the fan structure of a large-scale fan-spine system (see red and gray curves in \nfig{fig1}(e)). The top and side view of the inner magnetic null point system are shown in \nfig{fig1}(g) and (f), respectively, and the inner magnetic null point system is indicated by the black, cyan, pink, and orange curves. The height of the inner null point is measured to be about 17.4 Mm above the photosphere as indicated by the red arrow in \nfig{fig1}(f), and the height of the outer null is measured about 36 Mm. The green and yellow curves in the extrapolation results in \nfig{fig1}(e) -- (g) represent the two observed filaments F1 and F2, respectively. One can see that the two filaments are confined under the nested magnetic flux system.

\subsection{Eruption Details}
After checking AIA images carefully, we noticed that a sigmoidal, hot magnetic structure formed above the filaments (indicated by the green and yellow arrows in \nfig{fig2}(a)). The shape of the structure is indicated by the dashed white curves in \nfig{fig2}(b). The newly formed sigmoidal structure can only be seen in the high-temperature channels of the AIA (such as AIA 94 \AA\ and 131 \AA). Since it can not be observed in low-temperature wavelength channels of the AIA, the observed S-shaped structure should be a so-called hot channel \citep{2012NatCo...3..747Z}. To investigate the internal dynamics process between the newly formed hot channel and the large-scale fan-spine magnetic system, we made composite images by using the AIA 94 \AA, 131 \AA\ and 171 \AA\ running difference images (see \nfig{fig2}(d) -- (f)). We found that due to the gradual rising motion of the hot channel (indicated by the blue arrows in the locally enlarged images in \nfig{fig2}(d) -- (f)), the initial outer null point structure deformed into a thin and linear feature, as indicated by the red arrows in the locally enlarged images in \nfig{fig2}(d) and (e)). The appearance of the linear feature may suggest that the null point became increasingly compressed to form a BCS there. From 02:39 UT (Figure 2(e)) to 02:42 UT (Figure 2(f)), one can see that the BCS bent southwards, thinned, lengthened and was pushed away from the previous site to a certain distance (more details can be found in the online animation associated with \nfig{fig2}). As shown in the previous simulation studies \citep[e.g.,][]{2017Natur.544..452W, 2018ApJ...852...98W, 2021ApJ...909...54W}, such bend motion of BCS was closely related to the rising of an MFR. In the breakout model, the rising MFR can cause its overlying restraining field loops to expand and create a BCS at the separatrix between the closed bipolar flux system and the surrounding open or closed flux system. When the rising twisted flux forces against the separatrix, the explosive reconnection will occur and the reconnection site may be elongated and bent along the dome and, simultaneously, push the current sheet farther or higher. Therefore, comparing the phenomena in this event with the previous simulation results mentioned above, after 02:39 UT, the BCS was significantly bent and became longer and farther. This may represent that the hot channel front reached the breakout sheet and led to an explosive energy release. After the explosive reconnection, the hot channel erupted successfully, forming the obvious bright flare arcade below it (see \nfig{fig2}(c)). Then, a CME appeared in the FOVs of the LASCO/C2 and C3 (see \nfig{fig2}(g) and (h)). The dashed black curve and the white arrow in \nfig{fig2}(h) indicate the CME's bright front and core, respectively.

\begin{figure*}
\epsscale{0.85}
\figurenum{3}
\plotone{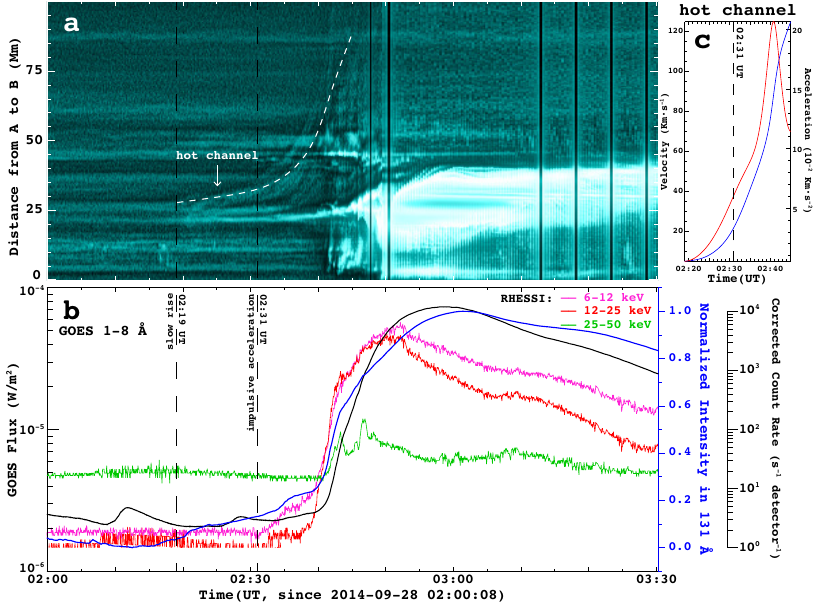}
\caption{Temporal evolution of the hot channel and synchronous fluxes. Panel (a) shows a time-distance diagram of AIA 131 \AA\ images made along the white line in \nfig{fig2}(b) (i.e., the white line in \nfig{fig4}(b)) enhanced by the multiscale gaussian normalization (MGN) method \citep{2014SoPh..289.2945M}. The slow rise and impulsive acceleration phase of the hot channel (indicated by a dashed white curve) have been observed in this time-distance diagram. Panel (b) shows {\em GOES} SXR fluxes in 1 -- 8 \AA\ (indicated by the black curve), the normalized light curve of AIA 131 \AA\ (denoted by the blue curve) obtained by integrating the region in the white box in \nfig{fig2}(b) and {\em RHESSI} X-ray fluxes at different energies in different colored curves. The speed (blue) and acceleration (red) of the hot channel in panel (c) were measured along the dashed white curve in panel (a). The left and right dashed vertical lines in panels (a) and (b) indicate the start time of the slow rise and the impulsive acceleration phase of the hot channel, respectively. 
\label{fig3}}
\end{figure*}

To further demonstrate the evolution of the hot channel in detail, we made a time-distance diagram using the AIA 131 \AA\ time-sequenced images along a path indicated by the white line in \nfig{fig2}(b). We plot the results in \nfig{fig3}(a). In the time-distance diagram (see \nfig{fig3}(a)), the hot channel first appeared at 02:19 UT (see the dashed white curve), and then, it showed two distinct eruption phases. The hot channel first experienced a slow-rising phase from 02:19 UT to 02:31 UT and then went through an impulsive acceleration phase starting from 02:31 UT (see the dashed white curve in \nfig{fig3}(a)). Such a kinematic behavior of an erupting MFR has been reported in many studies \citep[e.g.,][]{2012ApJ...750...12S, 2012NatCo...3..747Z, 2014ApJ...797L..15C, 2014ApJ...789...93C}. In the present event, the eruption speed and acceleration of the hot channel are plotted in \nfig{fig3}(c). During the slow-rising phase, the average speed is about \speed{9.9}, and the average acceleration is only about \accel{0.02}. In the impulsive acceleration phase, the speed of the hot channel increased quickly from about \speed{20.8} to \speed{124.9} within 14 minutes, with an average value of about \speed{70.7}. The acceleration of the hot channel also increased impulsively from about \accel{0.06} to about \accel{0.21}, with an average value of about \accel{0.14}, almost one order of magnitude larger than that during the slow-rising phase. Finally, the acceleration of the hot channel dropped rapidly from about 02:41 UT.

In the breakout model, an erupting flux rope often shows two distinct eruption phases \citep[e.g.,][]{2017Natur.544..452W, 2018ApJ...852...98W, 2018ApJ...854..155K}, whose erupting trajectory is similar to the depiction of the hot channel in \nfig{fig3}(a). In the breakout scenario, the dissipation of the current sheet under the rising magnetic flux by the so-called "tether-cutting" reconnection \citep{2001ApJ...552..833M} is commonly responsible for the formation of the erupting MFR. During this phase, the MFR would be formed and built up due to the continuous accumulation of the magnetic flux, and also rise slowly due to the magnetic buoyancy. The continuous rise of the MFR would lead to a positive feedback eruption process between the breakout reconnection and flare reconnection. In this scenario, breakout reconnection weakens the overlying strapped field lines, facilitating flare reconnection beneath the rising flux; the flare reconnection can further diminish the restraining flux above the rising flux, thereby promoting the breakout reconnection. The mutual facilitation between the breakout reconnection and flare reconnection proceeds in a positive feedback way. This can cause the motion of the MFR to convert from a slow rise to a fast rise. After this, the explosive reconnection may occur when the MFR front reaches the breakout sheet. These would be the reasons behind the dynamic evolution of two distinct eruption phases of an MFR in the breakout scenario. Based on such reasons, we here infer that the slow-rising phase of the observed hot channel may represent the formation and build-up phase of the erupting flux in the magnetic breakout model. The impulsive acceleration of the observed hot channel may include two parts according to the magnetic breakout scenario, i.e., the breakout feedback phase and the explosive reconnection process.
	
The above interpretation can be supported by the measurements of the X-ray fluxes at different energy channels and the light curves (see \nfig{fig3}(b)). Magnetic reconnection responsible for the MFR formation occurs beneath the core strapping field. During this phase, the magnetic reconnection should be moderate, leading to limited energy release. This may result in the build-up and slow-rising of the hot channel (02:19 UT -- 02:31 UT in \nfig{fig3}(a)) and a moderate increase in the light curve of AIA 131 \AA\ from 02:19 UT to 02:31 UT (see \nfig{fig3}(b)), whose integrated region was denoted by the white box in \nfig{fig2}(b). Following this, the feedback process between the breakout reconnection and flare reconnection would intensify the energy release, leading to a rapid rise of the hot channel (indicated by the dashed white curve after 02:31 UT in \nfig{fig3}(a)). Concurrently, the {\em RHESSI} flux curves at relatively low energy channels of 6 -- 12 keV and 12 -- 25 keV exhibited a significant increase. During the impulsive acceleration phase of the hot channel (after 02:31 UT), the {\em RHESSI} flux at relatively low energy channels of 6 -- 12 keV and 12 -- 25 keV also showed a transition at around 02:39 UT. Initially, from about 02:31 UT to 02:39 UT, they rose slowly, but after about 02:39 UT, the SXR and HXR flux curves all soared. This suggests a more significant increase in energy release around 02:39 UT. The feedback process is expected to accelerate the explosive breakout reconnection process when the hot channel front collides with the breakout sheet and then releases intensive energy. Notably, the transition of the flux curves during the impulsive acceleration at about 02:39 UT corresponds to the start time of the observed obvious deformation of the breakout sheet, as shown in \nfig{fig2}(e) and (f). This deformation would be regarded as an observational indicator of the hot channel front colliding with the breakout sheet (the relevant analysis has been discussed above). Therefore, the soaring flux curves (after 02:39 UT) likely relate to the explosive reconnection, aligning with the predictions in the breakout model \citep[e.g.,][]{2017Natur.544..452W, 2018ApJ...852...98W}.

\begin{figure*}
\epsscale{0.85}
\figurenum{4}
\plotone{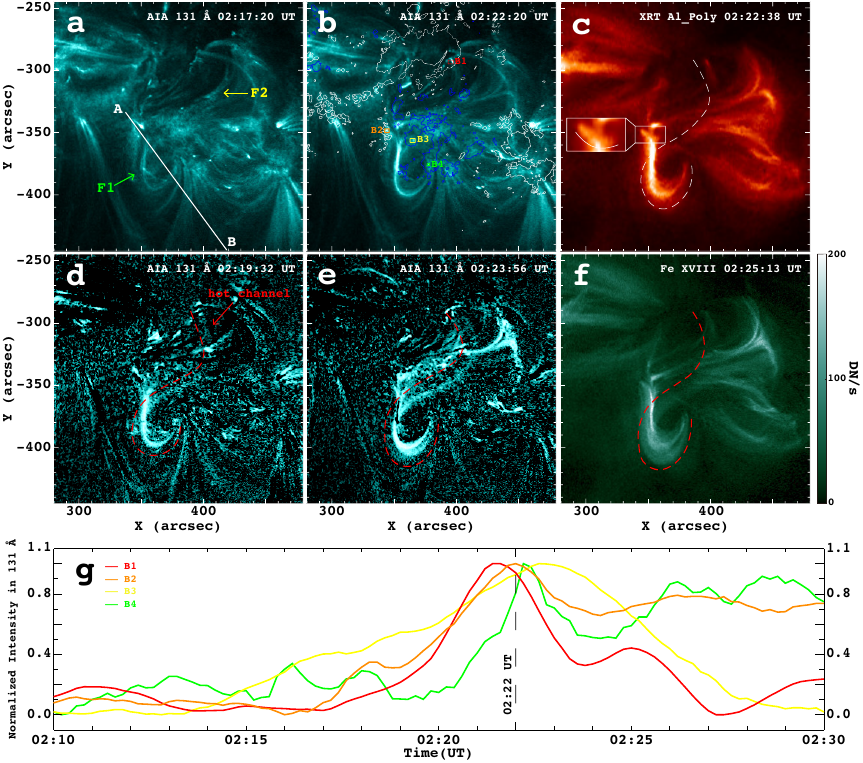}
\caption{Evolution of the source region near the newly formed MFR on 2014 September 28. Panels (a) -- (f) show the evolution of the source region displayed by AIA 131 \AA, XRT Al\_Poly and Fe \uppercase\expandafter{\romannumeral18} 93.932 \AA. F1 and F2 are denoted by the green and yellow arrows in panel (a), respectively. The white line AB in panel (a) denotes the path of the time-distance plot in \nfig{fig3}(a). In panel (b), white and blue contours refer to positive (300 G) and negative (-300 G) polarities, respectively. Panel (c) is XRT Al\_Poly image, and the white curve in enlarged box denotes a hot loop. Panels (d) and (e) are running difference images of AIA 131 \AA\ that have been enhanced via the gamma correction. The dashed curves in panels (c) -- (f) mark the axis of the newly formed hot channel MFR, which is shifting to the left to avoid blocking the MFR. In panel (g), the normalized light curves were obtained from AIA 131 \AA\ images. These curves have different colors corresponding to the boxes of the same color in panel (b). The vertical dashed line in panel (g) indicates the time when the brightenings in the boxes B1 -- B4 reached their peak. The animation covers a duration of about 47c minutes from 02:00 UT to 02:47 UT on 2014 September 28.
\\
(An animation of this figure is available)
\label{fig4}}
\end{figure*}

\begin{figure*}
\epsscale{0.85}
\figurenum{5}
\plotone{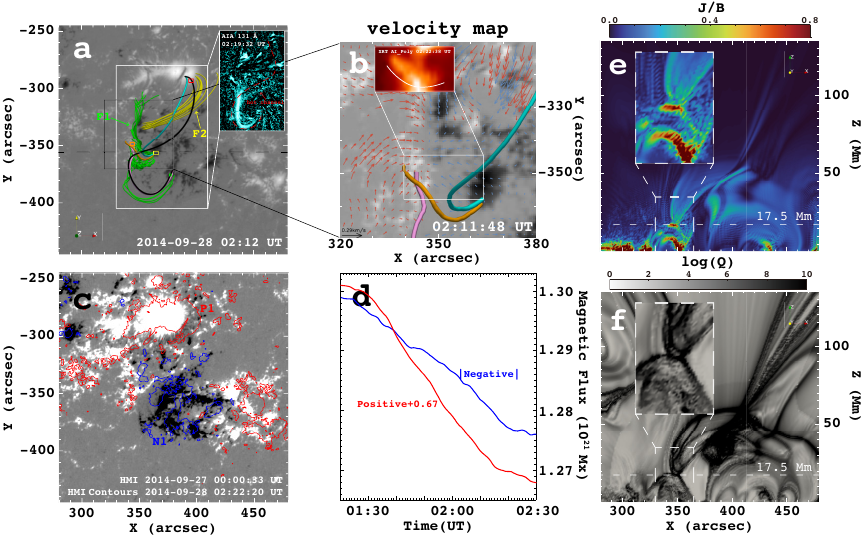}
\caption{The reconstructed 3D magnetic topology of the source region and relevant QSLs and $J/B$ distribution. Panel (a) is the top view of the inner magnetic null point system and has the same range with the black box in \nfig{fig1}(e). The green and yellow curves denote the observed two filaments. The enlarged plot in panel (a) shows the observed hot channel and displays the same region with the white box in panel (a). The red, orange, yellow and green boxes in panel (a) correspond to the colored boxes in \nfig{fig4}(b). Panel (b) shows the photospheric horizontal velocity map, determined by the FLCT method, overlaid with the extrapolated field lines. The enlarged plot in panel (b) is XRT Al\_Poly image at 02:22:38 UT, which contains a hot loop. Panels (e) and (f) show the distribution of the $J/B$ and QSLs along the dashed black line in panel (a), crossing the position where the inner null point exists at 02:12 UT on 2014 September 28. The dashed white lines in panels (e) and (f) indicate the height of the X-type distribution. Panel (c) is HMI image at 00:00:33 UT on 2014 September 27, overlaid with red (300 G; positive) and blue (-300 G; negative) contours at 02:22:20 UT on 2014 September 28. Panel (d) is the temporal evolution of the positive (red) and negative (blue) magnetic fluxes calculated over the same range with panel (b), from 01:20 UT to 02:30 UT on 2014 September 28. 
\label{fig5}}
\end{figure*}

\subsection{Activities in the Source Region}
\label{sec:33}
By observing the high-temperature channel videos made from the AIA 131 \AA, Fe \uppercase\expandafter{\romannumeral18} 93.932 \AA\ (removed the cool emission from AIA 94 \AA) and XRT Al\_Poly images of the source region (the animation of the \nfig{fig4} is available in the online journal for obtaining more details), we noticed that the inverted S-shaped hot channel mentioned above gradually appeared in AIA 131 \AA, the XRT Al\_Poly and Fe \uppercase\expandafter{\romannumeral18} 93.932 \AA\ images (see \nfig{fig4}(a) -- (f)), which indicates the formation of a hot MFR in this source region. Notably, a hot loop appeared in the XRT Al\_Poly image (see the insert plot in \nfig{fig4}(c)) shortly after the appearance of the hot channel. Since the hot channel was very faint during the initial formation stage, we used the 131 \AA\ running difference images enhanced via the gamma correction and the image obtained by removing the cool emission from AIA 94 \AA\ \citep{2013A&A...558A..73D}, only displaying the hotter Fe \uppercase\expandafter{\romannumeral18} 93.932 \AA\ (log $T$(K) $\sim$ 6.8) emission to better show the hot channel (see \nfig{fig4}(d) -- (f)). The profile of the hot channel was indicated by the dashed curves in \nfig{fig4} around the observed hot channel to guide eyes. During the slow rising phase of the hot channel (02:19 UT -- 02:31 UT), the hot channel became more and more evident in observations (see \nfig{fig4}(d) -- (f)) and the AIA 131 \AA\ light curve involving the hot channel also shows a slow increase trend (see blue curve in \nfig{fig3}(b)) from 02:19 UT to 02:31 UT, which may be the evidence for the formation and build-up processes of the hot channel during the slow rise phase.
	
We also observed four brightenings at around the footpoints of two filaments (see boxes B1 -- B4 in \nfig{fig4}(b)). By comparing \nfig{fig4}(b) with (c) -- (f), we can see that the two footpoints of the hot channel were rooted in opposite magnetic polarities close to the edges of the two magnetic poles of the active region (see B1 and B4 in \nfig{fig4}(b)). The normalized light curves within the colored box regions in \nfig{fig4}(b) were plotted in \nfig{fig4}(g). We noticed that the light curves increased fast at around 02:19 UT and reached their peak values at around 02:22 UT after the appearance of the hot channel (at around 02:19 UT). These suggest that the hot channel and four brightening patches might be related to each other temporally and spatially. Such close temporal and spatial relationships might hint at an intrinsic physical relationship among them. The potential physical relationships among them will be discussed in detail in Section 4.1.

\subsection{Magnetic Structure of the Source Region}
We reconstructed the pre-eruption magnetic configuration of the source region, and \nfig{fig5}(a) shows the top view of the inner magnetic null point system having the same range of the black box in \nfig{fig1}(e). The inner magnetic null point system existed under a large-scale fan-spine magnetic system (see the black, cyan, pink, and orange curves in \nfig{fig5}(a)). The green and yellow curves in the extrapolation results in \nfig{fig5}(a) were F1 and F2 (e.g., as noted by green and yellow arrows in \nfig{fig1}(a) and (d)). 

Based on the extrapolated magnetic fields, we calculated the current density (see \nfig{fig5}(e)) and the $Q$ factor (squashing degree, see \nfig{fig5}(f)) in the vertical plane passing through the magnetic null point of the inner magnetic null point system along the dashed black line indicated in \nfig{fig5}(a). The position of a magnetic null point is usually expected to be associated with a high current density and a high-degree $Q$ number. In principle, regions characterized by high $Q$ values are often liable to occur magnetic reconnection. By analyzing the current density and the $Q$ value in the current density and $Q$ maps, we can find that before the formation of the hot channel, X-type distributions of the current density and the $Q$ factor were already formed at about 02:12 UT at the position of the inner null point (see the insets in \nfig{fig5}(e) and (f)). In addition, based on these calculated current density and $Q$ maps, we can also answer whether the observed hot MFR was pre-existing in the corona. Because an MFR is a cylindrical current structure, one should identify a circular structure representing the MFR boundary in the current density and $Q$ maps perpendicular to the axis of the MFR \citep[e.g.,][]{2012ApJ...750...15S,2013ApJ...779..157G,2016ApJ...818..148L}. In the present case, we do not observe such a structure before the presence of the hot channel in observations. Therefore, the observed hot channel might not be pre-existing. 
	
\section{Conclusions and Discussions}
\label{sec:4}  
In this paper, we present the observations of a CME originating from a complicated magnetic source region on 2014 September 28 by using multi-wavelength imaging observations taken by the {\em SDO}, the NVST, the {\em Hinode}, and the {\em SOHO}, and X-ray fluxes recorded by the {\em GOES} and the {\em RHESSI} satellites combined with the NLFFF extrapolation. Our analysis results indicate that the magnetic structure in the eruption source region is a nested double null magnetic system in which a large-scale fan-spine magnetic system hosts a small-scale magnetic null point system below. The outer fan-spine magnetic system hosting a magnetic null can be observed directly in EUV images taken by the AIA. In contrast, the inner magnetic null point system is indirectly revealed by the three-dimensional coronal magnetic field. Although, there were two filaments in the eruption source region, they kept stable during the whole eruption process. Our analysis results indicate that the observed flare and CME were probably caused by the eruption of a newly formed hot channel rather than the observed filaments. In addition, the observed circular flare ribbon, remote ribbon, and the two conjugated flare ribbons underneath the erupting hot channel evidenced the interaction process between the newly formed MFR and the outer fan-spine magnetic system. The present study reveals that the formation of a CME is more complicated than we thought and how an MFR forms and erupts as a large-scale CME from such a complex nested double null magnetic system. The formation of the hot channel, the reasons why the observed filaments did not erupt, and the formation of the CME in the nested double null magnetic system will be discussed in the following sections.

\subsection{Formation Mechanism of the Hot Channel MFR}
Based on the extrapolated three-dimensional coronal magnetic field, we find an inner magnetic null point system nested by an outer fan-spine magnetic system (see \nfig{fig1}). In the inner magnetic null point system, if the magnetic reconnection occurred between the pink and cyan curves, the loops analogous to the black one can be formed. \nfig{fig5}(a) shows that the location and shape of the black loop overlying the inner null point is similar to the observed hot channel, i.e., they are both in the shape of an inverse sigmoid and share nearly the same pair of the footpoints. Such similarities suggest that the formation of the observed hot channel can be due to the magnetic reconnection between the pink and cyan field lines at the inner null point. According to the extrapolation results, such a magnetic reconnection could lead to brightenings at the footpoints of the magnetic field lines consisting of the inner magnetic null point system. The imaging observation well showed four brightenings whose locations are spatially consistent with the four footpoints of the extrapolated magnetic field lines comprised of the inner magnetic null point system (see \nfig{fig4}(b) and \nfig{fig5}(a)), which suggest that magnetic reconnection did occur at the inner null point. Besides the magnetic reconnection between the pink and cyan field lines, the same brightenings can also be created by the magnetic reconnection between the orange and black field lines.By checking the XRT Al\_Poly soft X-ray images, we observed a hot loop closely following the hot channel’s appearance at about 02:22:38 UT. This hot loop was located under the inner null point, and it had a similar shape and footpoint locations to the extrapolated orange curve (see \nfig{fig5}(b)). In addition, this hot loop’s appearance time was close to the peak time of B2 and B3. Based on these close temporal and spatial relationships between the extrapolated magnetic field lines and the actual observations, we conclude that the magnetic reconnection at the inner null point should occur between the extrapolated pink and cyan field lines, where the observed hot loop in the XRT Al\_Poly images and the hot channel MFR in EUV and soft X-ray images represent the downward and upward moving reconnected field lines, respectively.

To investigate the trigger of such a magnetic reconnection, we further checked the velocity map of the LOS magnetic field using the Fourier local correlation tracking (FLCT) method \citep{2008ASPC..383..373F}. The region of interest of the velocity map is around the PIL focusing on the inner magnetic null point system (the black box in \nfig{fig5}(a)), and the time of the velocity map is before the formation of the hot channel at 02:11:48 UT. The calculated velocity map is superimposed on the LOS magnetogram at 02:11:48 UT as red and blue arrows in \nfig{fig5}(b). At the same time, the colored curves are the representative extrapolated magnetic field lines the same as those in Figure 5(a). One can see that the magnetic polarities show an apparent clockwise swirling motion. Due to the high $\beta$ nature of the photosphere, the magnetic field lines are frozen in the photospheric flow and migrate forced by the photosphere flow. Such photospheric swirling motion might lead to the approaching of magnetic field lines rooted in the positive and negative polarities, such as the pink and cyan ones, and therefore trigger magnetic reconnection at the inner null point. The photospheric swirling motion lasted for about one hour before the eruption from 01:10 UT to 02:20 UT, and the free energy generated by the swirling motion is estimated at about $9.75$ $\times$ $10^{30}$ $erg$ (see appendix for more details). Meanwhile, the positive and negative magnetic fluxes calculated over the region of interest both show a continuous decreasing trend (see \nfig{fig5}(d)). This decrease in the magnetic fluxes is suggestive of the flux cancellation driven by the magnetic reconnection \citep[e.g.,][]{1985AuJPh..38..929M, 1985AuJPh..38..855L, 1999ApJ...515..435L, 2018ApJ...862L..24P}. The velocity map also revealed that there existed the component of the photospheric converging motion at the source region. Such converging motion is in favor of the occurrence of the magnetic reconnection. The total magnetic energy released during flux cancellation is $E =$ $\frac{1}{8\pi}$ $B$$\Phi$$L$ \citep{2020A&A...644A.130C}, where $B$ is the magnetic field strength, $\Phi$ is total cancelled magnetic flux, and $L$ is the length of the current sheet. We assume that during the flux cancellation period, $\Phi$ $\approx$ $0.04$ $\times$ $10^{21}$ Mx by calculating the balance part of the flux loss (see \nfig{fig5}(d)), and $B$ $\approx$ $300$ $G$. Based on the extrapolated results, we simply assume that the current sheet is confined by the upper black and lower orange loops, meaning $L$ $\approx$ $6.7$ Mm. Thus, the released magnetic energy is about $3.2$ $\times$ $10^{29}$ $erg$, which would be sufficient to power a B-class flare in the solar corona \citep{2011SSRv..159..263H}. Note that some of the loops generated by the magnetic reconnection might submerge and act as magnetic cancellations, but others might still stay in the atmosphere. The loops left in the atmosphere could continue to reconnect, further releasing magnetic energy. This part can not be included by the simple quantitative calculation ($\frac{1}{8\pi}$ $B$$\Phi$$L$), which could lead to the underestimate of the released magnetic energy. Thus, in reality, the released magnetic energy might be greater than the estimated value by the simple assumption. In addition, we checked the HMI images for over 26 hours from 00:00:33 UT on 2014 September 27 to 02:22:20 UT on 2014 September 28 (see \nfig{fig5}(c)). Notably, the two main regions of opposite magnetic polarities (P1 and N1; noted in \nfig{fig5}(c)) sheared significantly. The cyan and pink filed lines in \nfig{fig5}(a) each have one footpoint rooted at P1 and N1, respectively. Such shearing motion between P1 and N1 might inject free energy into this source region throughout more than 1 day, which might generate enough energy to power the eruption. Hence, the photospheric swirl motions revealed by the velocity map in \nfig{fig5}(b) might trigger reconnection in the inner magnetic null point system, releasing the free energy built up by the footpoint shearing and contributing to the formation and build-up of the hot channel.

\subsection{Stability of the Two Filaments in the Source Region}
The two filaments located under the inner magnetic null point system in the eruption source region seemed detached from the whole eruption process and remained stable at their original sites (see \nfig{fig1}(d)). In the breakout model, a filament eruption is not required. It depends on where the filament is located relative to the flare reconnection site \citep[e.g.,][]{2021ApJ...907...41K}. The erupting MFR is a newly formed hot channel in the present study, which was formed within the inner magnetic null point system over the two filaments. As the magnetic reconnection occurs in the inner magnetic null point system, the magnetic field lines will be rearranged into newly formed confining loops above the filaments. As the MFR continuously rises, flare reconnection will occur underneath the MFR, and more new magnetic flux will accumulate in the MFR. At the same time, the flare arcade created by the flare reconnection between the two legs of the confining field lines of the MFR will further strengthen the confinement above the filaments. All these processes occurred will effectively ensure the stability of the filaments.

\subsection{Cause of the MFR's Successful Eruption}
The magnetic breakout configuration widely exists in the solar atmosphere and it has been confirmed by many observational and numerical studies \citep[e.g.,][]{1999ApJ...510..485A, 2004ApJ...614.1028M, 2008ApJ...680..740D, 2012ApJ...750...12S, 2013ApJ...764...87L, 2016ApJ...820L..37C}. Previous studies also suggested that a small flux system can be nested into a larger one, which can make up an overall nested double null magnetic system, and eruptions from it might be more complex \citep[e.g.,][]{2019ApJ...871....4H, 2019ApJ...886L..34L, 2024ApJ...966...27K}. \cite{2024ApJ...966...27K} recently studied a nested fan-spine magnetic system in which a fan-spine magnetic system is nested by a large transequatorial pseudo-streamer containing a null. The eruption of a flux rope from the inner fan-spine system did not cause a large-scale CME in the outer corona except for a weak shock and a faint collimated outflow from the pseudo-streamer. The nested fan-spine magnetic system hosts two null points. Therefore, it can also be called a nested double null magnetic system. Combining observation and simulation, \cite{2024ApJ...966...27K} conducted a systematic study on how the nested double null magnetic system modulates the flux rope eruption within the system. The authors concluded that the nested double null magnetic system can significantly restrict the eruption of the MFR from the inner fan-spine system, which can account for the observed weak jet and stalled ejection. Although the basic magnetic topology in the present event is similar to that in \cite{2024ApJ...966...27K}, i.e., a nested double null magnetic system, the MFR eruption successfully caused a large-scale CME in the outer corona in our event. To determine what causes such a significant difference between the two events, we would like to compare our event to the one reported in \cite{2024ApJ...966...27K}. 

The nested double null magnetic system in \cite{2024ApJ...966...27K} can be divided into three flux systems by the two null points, i.e., the lower part contains the rising flux rope; the middle part acts as a strapping field that is anti-parallel to the lower and upper fluxes; and the upper part also acts as a strapping field but is parallel to the lower one. Due to the rise of the MFR, the magnetic null points will be squeezed into current sheets, and breakout reconnection within them will dissipate the middle strapping field from both directions. Since the flux rope field is anti-parallel to the the middle flux, its twist and shear will be dispersed within the system when it reaches the middle part strapping field. After reaching the outer null, the downward magnetic tension force caused by the upper trapping flux will prevent the rising of the MFR, due to the same orientation between the upper strapping flux and the toroidal field component of the rising flux. In this process, the upper confining field acts like a wall that hinders the eruption of the rising flux, similar to the effect of a slingshot. Hence, the nested double null magnetic system can effectively restrict the eruption of the MFR from the inner fan-spine system and, therefore, naturally results in weak or stalled ejections as observed in \cite{2024ApJ...966...27K}. In the present event, it is clear that the newly formed MFR is above the inner null point but below the outer one. This is the same as the single null point breakout topology for the MFR. Therefore, the MFR eruption should proceed similarly to the single null point breakout eruption, in which the rising MFR first triggers the breakout reconnection at the outer null point to open the external strapping magnetic flux, and then the flare reconnection underneath the MFR drives the violent eruption of the MFR to produce the CME. 

In comparison to \cite{2024ApJ...966...27K}, the main difference is the initial birth location of the MFR relative to the inner null point, which may determine whether an MFR can erupt successfully or not in the nested double null magnetic system. Besides the formation position of the erupting flux, the kink or rotation motion of the flux rope and the amount of the overlying strapping field above the rising MFR via external/internal reconnection may also affect the eruption process of the MFR. Thus, solar eruptions from the nested double null magnetic system deserve further relevant theoretical and observational investigations.

\begin{acknowledgments}
We thank the {\em SDO}, the NVST, the {\em RHESSI}, the {\em GOES}, the {\em Hinode}, and the {\em SOHO} teams for data support, and the anonymous referee who provided many valuable comments to improve quality of this paper. This work is supported by the Natural Science Foundation of China (12173083, 12303062, 11922307) and the Sichuan Science and Technology Program (2023NSFSC1351).
\end{acknowledgments}

\begin{appendix}
\section*{Calculation of the Poynting Flux}
In this paper, we estimate the Poynting flux generated by the swirling motions using the formula $F = -$ $\frac{1}{4\pi}$ $B_v$ \bm{$B_h$} $\cdot$ \bm{$V_h$}, where $B_v$ and \bm{$B_h$} are the vertical and horizontal components of the field and \bm{$V_h$} is the horizontal velocity of the footpoint \citep{2006SoPh..234...41K}. We assume $B_v$ $=$ $B_h$ $\approx$ $300$ $G$ based on the observations, and $V_h$ $\approx$ \speed{0.2}. Thus, the estimated Poynting flux is $1.432$ $\times$ $10^{8}$ $erg$ $cm^{-2}$ $s^{-1}$. By taking the swirling area $S$ $\approx$ $1.892$ $\times$ $10^{19}$ $cm^2$ ($60\arcsec$ $\times$ $60\arcsec$; see \nfig{fig5}(b)) and the duration of the swirl motions (almost an hour) into account, the free energy generated by the swirling motions is estimated about $9.75$ $\times$ $10^{30}$ $erg$. The Poynting flux driven by the photosoheric swirling motions can afford enough energy that is required by the buildup process of the hot channel.
\end{appendix}
 

\begin{thebibliography}{}
\expandafter\ifx\csname natexlab\endcsname\relax\def\natexlab#1{#1}\fi
\providecommand{\url}[1]{\href{#1}{#1}}
\providecommand{\dodoi}[1]{doi:~\href{http://doi.org/#1}{\nolinkurl{#1}}}
\providecommand{\doeprint}[1]{\href{http://ascl.net/#1}{\nolinkurl{http://ascl.net/#1}}}
\providecommand{\doarXiv}[1]{\href{https://arxiv.org/abs/#1}{\nolinkurl{https://arxiv.org/abs/#1}}}

\bibitem[{{Antiochos} {et~al.}(1999){Antiochos}, {DeVore}, \&
  {Klimchuk}}]{1999ApJ...510..485A}
{Antiochos}, S.~K., {DeVore}, C.~R., \& {Klimchuk}, J.~A. 1999, \apj, 510, 485,
  \dodoi{10.1086/306563}

\bibitem[{{Aulanier} {et~al.}(1998){Aulanier}, {Demoulin}, {van
  Driel-Gesztelyi}, {Mein}, \& {Deforest}}]{1998A&A...335..309A}
{Aulanier}, G., {Demoulin}, P., {van Driel-Gesztelyi}, L., {Mein}, P., \&
  {Deforest}, C. 1998, \aap, 335, 309

\bibitem[{{Aulanier} {et~al.}(2010){Aulanier}, {T{\"o}r{\"o}k}, {D{\'e}moulin},
  \& {DeLuca}}]{2010ApJ...708..314A}
{Aulanier}, G., {T{\"o}r{\"o}k}, T., {D{\'e}moulin}, P., \& {DeLuca}, E.~E.
  2010, \apj, 708, 314, \dodoi{10.1088/0004-637X/708/1/314}

\bibitem[{{Brueckner} {et~al.}(1995){Brueckner}, {Howard}, {Koomen},
  {Korendyke}, {Michels}, {Moses}, {Socker}, {Dere}, {Lamy}, {Llebaria},
  {Bout}, {Schwenn}, {Simnett}, {Bedford}, \& {Eyles}}]{1995SoPh..162..357B}
{Brueckner}, G.~E., {Howard}, R.~A., {Koomen}, M.~J., {et~al.} 1995, \solphys,
  162, 357, \dodoi{10.1007/BF00733434}

\bibitem[{{Carmichael}(1964)}]{1964NASSP..50..451C}
{Carmichael}, H. 1964, in NASA Special Publication, Vol.~50, 451

\bibitem[{{Chen} {et~al.}(2014){Chen}, {Zhang}, {Cheng}, {Ma}, {Yang}, \&
  {Li}}]{2014ApJ...797L..15C}
{Chen}, H., {Zhang}, J., {Cheng}, X., {et~al.} 2014, \apjl, 797, L15,
  \dodoi{10.1088/2041-8205/797/2/L15}

\bibitem[{{Chen} {et~al.}(2016){Chen}, {Du}, {Zhao}, {Wu}, {Liu}, {Wang},
  {Ruan}, {Feng}, \& {Song}}]{2016ApJ...820L..37C}
{Chen}, Y., {Du}, G., {Zhao}, D., {et~al.} 2016, \apjl, 820, L37,
  \dodoi{10.3847/2041-8205/820/2/L37}

\bibitem[{{Cheng} {et~al.}(2014){Cheng}, {Ding}, {Zhang}, {Sun}, {Guo}, {Wang},
  {Kliem}, \& {Deng}}]{2014ApJ...789...93C}
{Cheng}, X., {Ding}, M.~D., {Zhang}, J., {et~al.} 2014, \apj, 789, 93,
  \dodoi{10.1088/0004-637X/789/2/93}

\bibitem[{{Cheng} {et~al.}(2013){Cheng}, {Zhang}, {Ding}, {Liu}, \&
  {Poomvises}}]{2013ApJ...763...43C}
{Cheng}, X., {Zhang}, J., {Ding}, M.~D., {Liu}, Y., \& {Poomvises}, W. 2013,
  \apj, 763, 43, \dodoi{10.1088/0004-637X/763/1/43}

\bibitem[{{Cheng} {et~al.}(2011){Cheng}, {Zhang}, {Liu}, \&
  {Ding}}]{2011ApJ...732L..25C}
{Cheng}, X., {Zhang}, J., {Liu}, Y., \& {Ding}, M.~D. 2011, \apjl, 732, L25,
  \dodoi{10.1088/2041-8205/732/2/L25}

\bibitem[{{Chitta} {et~al.}(2020){Chitta}, {Peter}, {Priest}, \&
  {Solanki}}]{2020A&A...644A.130C}
{Chitta}, L.~P., {Peter}, H., {Priest}, E.~R., \& {Solanki}, S.~K. 2020, \aap,
  644, A130, \dodoi{10.1051/0004-6361/202039099}

\bibitem[{{Del Zanna}(2013)}]{2013A&A...558A..73D}
{Del Zanna}, G. 2013, \aap, 558, A73, \dodoi{10.1051/0004-6361/201321653}

\bibitem[{{DeVore} \& {Antiochos}(2008)}]{2008ApJ...680..740D}
{DeVore}, C.~R., \& {Antiochos}, S.~K. 2008, \apj, 680, 740,
  \dodoi{10.1086/588011}

\bibitem[{{Fisher} \& {Welsch}(2008)}]{2008ASPC..383..373F}
{Fisher}, G.~H., \& {Welsch}, B.~T. 2008, in Astronomical Society of the
  Pacific Conference Series, Vol. 383, Subsurface and Atmospheric Influences on
  Solar Activity, ed. R.~{Howe}, R.~W. {Komm}, K.~S. {Balasubramaniam}, \&
  G.~J.~D. {Petrie}, 373

\bibitem[{{Golub} {et~al.}(2007){Golub}, {DeLuca}, {Austin}, {Bookbinder},
  {Caldwell}, {Cheimets}, {Cirtain}, {Cosmo}, {Reid}, {Sette}, {Weber},
  {Sakao}, {Kano}, {Shibasaki}, {Hara}, {Tsuneta}, {Kumagai}, {Tamura},
  {Shimojo}, {McCracken}, {Carpenter}, {Haight}, {Siler}, {Wright}, {Tucker},
  {Rutledge}, {Barbera}, {Peres}, \& {Varisco}}]{2007SoPh..243...63G}
{Golub}, L., {DeLuca}, E., {Austin}, G., {et~al.} 2007, \solphys, 243, 63,
  \dodoi{10.1007/s11207-007-0182-1}

\bibitem[{{Guo} {et~al.}(2013){Guo}, {Ding}, {Cheng}, {Zhao}, \&
  {Pariat}}]{2013ApJ...779..157G}
{Guo}, Y., {Ding}, M.~D., {Cheng}, X., {Zhao}, J., \& {Pariat}, E. 2013, \apj,
  779, 157, \dodoi{10.1088/0004-637X/779/2/157}

\bibitem[{{Hannah} {et~al.}(2011){Hannah}, {Hudson}, {Battaglia}, {Christe},
  {Ka{\v{s}}parov{\'a}}, {Krucker}, {Kundu}, \&
  {Veronig}}]{2011SSRv..159..263H}
{Hannah}, I.~G., {Hudson}, H.~S., {Battaglia}, M., {et~al.} 2011, \ssr, 159,
  263, \dodoi{10.1007/s11214-010-9705-4}

\bibitem[{{Hirayama}(1974)}]{1974SoPh...34..323H}
{Hirayama}, T. 1974, \solphys, 34, 323, \dodoi{10.1007/BF00153671}

\bibitem[{{Hong} {et~al.}(2019){Hong}, {Yang}, {Chen}, {Bi}, {Yang}, \&
  {Chen}}]{2019ApJ...874..146H}
{Hong}, J., {Yang}, J., {Chen}, H., {et~al.} 2019, \apj, 874, 146,
  \dodoi{10.3847/1538-4357/ab0c9d}

\bibitem[{{Hou} {et~al.}(2019){Hou}, {Li}, {Yang}, \&
  {Zhang}}]{2019ApJ...871....4H}
{Hou}, Y., {Li}, T., {Yang}, S., \& {Zhang}, J. 2019, \apj, 871, 4,
  \dodoi{10.3847/1538-4357/aaf4f4}

\bibitem[{{Ji} {et~al.}(2003){Ji}, {Wang}, {Schmahl}, {Moon}, \&
  {Jiang}}]{2003ApJ...595L.135J}
{Ji}, H., {Wang}, H., {Schmahl}, E.~J., {Moon}, Y.~J., \& {Jiang}, Y. 2003,
  \apjl, 595, L135, \dodoi{10.1086/378178}

\bibitem[{{Karpen} {et~al.}(2012){Karpen}, {Antiochos}, \&
  {DeVore}}]{2012ApJ...760...81K}
{Karpen}, J.~T., {Antiochos}, S.~K., \& {DeVore}, C.~R. 2012, \apj, 760, 81,
  \dodoi{10.1088/0004-637X/760/1/81}

\bibitem[{{Karpen} {et~al.}(2024){Karpen}, {Kumar}, {Wyper}, {DeVore}, \&
  {Antiochos}}]{2024ApJ...966...27K}
{Karpen}, J.~T., {Kumar}, P., {Wyper}, P.~F., {DeVore}, C.~R., \& {Antiochos},
  S.~K. 2024, \apj, 966, 27, \dodoi{10.3847/1538-4357/ad2eaa}

\bibitem[{{Kliem} \& {T{\"o}r{\"o}k}(2006)}]{2006PhRvL..96y5002K}
{Kliem}, B., \& {T{\"o}r{\"o}k}, T. 2006, \prl, 96, 255002,
  \dodoi{10.1103/PhysRevLett.96.255002}

\bibitem[{{Klimchuk}(2006)}]{2006SoPh..234...41K}
{Klimchuk}, J.~A. 2006, \solphys, 234, 41, \dodoi{10.1007/s11207-006-0055-z}

\bibitem[{{Kopp} \& {Pneuman}(1976)}]{1976SoPh...50...85K}
{Kopp}, R.~A., \& {Pneuman}, G.~W. 1976, \solphys, 50, 85,
  \dodoi{10.1007/BF00206193}

\bibitem[{{Kosugi} {et~al.}(2007){Kosugi}, {Matsuzaki}, {Sakao}, {Shimizu},
  {Sone}, {Tachikawa}, {Hashimoto}, {Minesugi}, {Ohnishi}, {Yamada}, {Tsuneta},
  {Hara}, {Ichimoto}, {Suematsu}, {Shimojo}, {Watanabe}, {Shimada}, {Davis},
  {Hill}, {Owens}, {Title}, {Culhane}, {Harra}, {Doschek}, \&
  {Golub}}]{2007SoPh..243....3K}
{Kosugi}, T., {Matsuzaki}, K., {Sakao}, T., {et~al.} 2007, \solphys, 243, 3,
  \dodoi{10.1007/s11207-007-9014-6}

\bibitem[{{Kumar} {et~al.}(2018){Kumar}, {Karpen}, {Antiochos}, {Wyper},
  {DeVore}, \& {DeForest}}]{2018ApJ...854..155K}
{Kumar}, P., {Karpen}, J.~T., {Antiochos}, S.~K., {et~al.} 2018, \apj, 854,
  155, \dodoi{10.3847/1538-4357/aaab4f}

\bibitem[{{Kumar} {et~al.}(2021){Kumar}, {Karpen}, {Antiochos}, {Wyper},
  {DeVore}, \& {Lynch}}]{2021ApJ...907...41K}
---. 2021, \apj, 907, 41, \dodoi{10.3847/1538-4357/abca8b}

\bibitem[{{Lemen} {et~al.}(2012){Lemen}, {Title}, {Akin}, {Boerner}, {Chou},
  {Drake}, {Duncan}, {Edwards}, {Friedlaender}, {Heyman}, {Hurlburt}, {Katz},
  {Kushner}, {Levay}, {Lindgren}, {Mathur}, {McFeaters}, {Mitchell}, {Rehse},
  {Schrijver}, {Springer}, {Stern}, {Tarbell}, {Wuelser}, {Wolfson}, {Yanari},
  {Bookbinder}, {Cheimets}, {Caldwell}, {Deluca}, {Gates}, {Golub}, {Park},
  {Podgorski}, {Bush}, {Scherrer}, {Gummin}, {Smith}, {Auker}, {Jerram},
  {Pool}, {Soufli}, {Windt}, {Beardsley}, {Clapp}, {Lang}, \&
  {Waltham}}]{2012SoPh..275...17L}
{Lemen}, J.~R., {Title}, A.~M., {Akin}, D.~J., {et~al.} 2012, \solphys, 275,
  17, \dodoi{10.1007/s11207-011-9776-8}

\bibitem[{{Li} {et~al.}(2019){Li}, {Yang}, {Hong}, \&
  {Chen}}]{2019ApJ...886L..34L}
{Li}, H., {Yang}, J., {Hong}, J., \& {Chen}, H. 2019, \apjl, 886, L34,
  \dodoi{10.3847/2041-8213/ab564e}

\bibitem[{{Lin} \& {Forbes}(2000)}]{2000JGR...105.2375L}
{Lin}, J., \& {Forbes}, T.~G. 2000, \jgr, 105, 2375,
  \dodoi{10.1029/1999JA900477}

\bibitem[{{Lin} {et~al.}(2002){Lin}, {Dennis}, {Hurford}, {Smith}, {Zehnder},
  {Harvey}, {Curtis}, {Pankow}, {Turin}, {Bester}, {Csillaghy}, {Lewis},
  {Madden}, {van Beek}, {Appleby}, {Raudorf}, {McTiernan}, {Ramaty}, {Schmahl},
  {Schwartz}, {Krucker}, {Abiad}, {Quinn}, {Berg}, {Hashii}, {Sterling},
  {Jackson}, {Pratt}, {Campbell}, {Malone}, {Landis}, {Barrington-Leigh},
  {Slassi-Sennou}, {Cork}, {Clark}, {Amato}, {Orwig}, {Boyle}, {Banks},
  {Shirey}, {Tolbert}, {Zarro}, {Snow}, {Thomsen}, {Henneck}, {Mchedlishvili},
  {Ming}, {Fivian}, {Jordan}, {Wanner}, {Crubb}, {Preble}, {Matranga}, {Benz},
  {Hudson}, {Canfield}, {Holman}, {Crannell}, {Kosugi}, {Emslie}, {Vilmer},
  {Brown}, {Johns-Krull}, {Aschwanden}, {Metcalf}, \&
  {Conway}}]{2002SoPh..210....3L}
{Lin}, R.~P., {Dennis}, B.~R., {Hurford}, G.~J., {et~al.} 2002, \solphys, 210,
  3, \dodoi{10.1023/A:1022428818870}

\bibitem[{{Litvinenko}(1999)}]{1999ApJ...515..435L}
{Litvinenko}, Y.~E. 1999, \apj, 515, 435, \dodoi{10.1086/307001}

\bibitem[{{Liu}(2020)}]{2020RAA....20..165L}
{Liu}, R. 2020, Research in Astronomy and Astrophysics, 20, 165,
  \dodoi{10.1088/1674-4527/20/10/165}

\bibitem[{{Liu} {et~al.}(2016){Liu}, {Kliem}, {Titov}, {Chen}, {Wang}, {Wang},
  {Liu}, {Xu}, \& {Wiegelmann}}]{2016ApJ...818..148L}
{Liu}, R., {Kliem}, B., {Titov}, V.~S., {et~al.} 2016, \apj, 818, 148,
  \dodoi{10.3847/0004-637X/818/2/148}

\bibitem[{{Liu} {et~al.}(2014){Liu}, {Xu}, {Gu}, {Wang}, {You}, {Shen}, {Lu},
  {Jin}, {Chen}, {Lou}, {Li}, {Liu}, {Xu}, {Rao}, {Hu}, {Li}, {Fu}, {Wang},
  {Bao}, {Wu}, \& {Zhang}}]{2014RAA....14..705L}
{Liu}, Z., {Xu}, J., {Gu}, B.-Z., {et~al.} 2014, Research in Astronomy and
  Astrophysics, 14, 705, \dodoi{10.1088/1674-4527/14/6/009}

\bibitem[{{Livi} {et~al.}(1985){Livi}, {Wang}, \&
  {Martin}}]{1985AuJPh..38..855L}
{Livi}, S.~H.~B., {Wang}, J., \& {Martin}, S.~F. 1985, Australian Journal of
  Physics, 38, 855, \dodoi{10.1071/PH850855}

\bibitem[{{Lynch} \& {Edmondson}(2013)}]{2013ApJ...764...87L}
{Lynch}, B.~J., \& {Edmondson}, J.~K. 2013, \apj, 764, 87,
  \dodoi{10.1088/0004-637X/764/1/87}

\bibitem[{{Mackay} {et~al.}(2010){Mackay}, {Karpen}, {Ballester}, {Schmieder},
  \& {Aulanier}}]{2010SSRv..151..333M}
{Mackay}, D.~H., {Karpen}, J.~T., {Ballester}, J.~L., {Schmieder}, B., \&
  {Aulanier}, G. 2010, \ssr, 151, 333, \dodoi{10.1007/s11214-010-9628-0}

\bibitem[{{MacNeice} {et~al.}(2004){MacNeice}, {Antiochos}, {Phillips},
  {Spicer}, {DeVore}, \& {Olson}}]{2004ApJ...614.1028M}
{MacNeice}, P., {Antiochos}, S.~K., {Phillips}, A., {et~al.} 2004, \apj, 614,
  1028, \dodoi{10.1086/423887}

\bibitem[{{Martin} {et~al.}(1985){Martin}, {Livi}, \&
  {Wang}}]{1985AuJPh..38..929M}
{Martin}, S.~F., {Livi}, S.~H.~B., \& {Wang}, J. 1985, Australian Journal of
  Physics, 38, 929, \dodoi{10.1071/PH850929}

\bibitem[{{Moore} {et~al.}(2001){Moore}, {Sterling}, {Hudson}, \&
  {Lemen}}]{2001ApJ...552..833M}
{Moore}, R.~L., {Sterling}, A.~C., {Hudson}, H.~S., \& {Lemen}, J.~R. 2001,
  \apj, 552, 833, \dodoi{10.1086/320559}

\bibitem[{{Morgan} \& {Druckm{\"u}ller}(2014)}]{2014SoPh..289.2945M}
{Morgan}, H., \& {Druckm{\"u}ller}, M. 2014, \solphys, 289, 2945,
  \dodoi{10.1007/s11207-014-0523-9}

\bibitem[{{M{\"o}stl} {et~al.}(2018){M{\"o}stl}, {Amerstorfer}, {Palmerio},
  {Isavnin}, {Farrugia}, {Lowder}, {Winslow}, {Donnerer}, {Kilpua}, \&
  {Boakes}}]{2018SpWea..16..216M}
{M{\"o}stl}, C., {Amerstorfer}, T., {Palmerio}, E., {et~al.} 2018, Space
  Weather, 16, 216, \dodoi{10.1002/2017SW001735}

\bibitem[{{Pariat} \& {D{\'e}moulin}(2012)}]{2012A&A...541A..78P}
{Pariat}, E., \& {D{\'e}moulin}, P. 2012, \aap, 541, A78,
  \dodoi{10.1051/0004-6361/201118515}

\bibitem[{{Pesnell} {et~al.}(2012){Pesnell}, {Thompson}, \&
  {Chamberlin}}]{2012SoPh..275....3P}
{Pesnell}, W.~D., {Thompson}, B.~J., \& {Chamberlin}, P.~C. 2012, \solphys,
  275, 3, \dodoi{10.1007/s11207-011-9841-3}

\bibitem[{{Priest}(2014)}]{2014masu.book.....P}
{Priest}, E. 2014, {Magnetohydrodynamics of the Sun},
  \dodoi{10.1017/CBO9781139020732}

\bibitem[{{Priest} {et~al.}(2018){Priest}, {Chitta}, \&
  {Syntelis}}]{2018ApJ...862L..24P}
{Priest}, E.~R., {Chitta}, L.~P., \& {Syntelis}, P. 2018, \apjl, 862, L24,
  \dodoi{10.3847/2041-8213/aad4fc}

\bibitem[{{Priest} \& {D{\'e}moulin}(1995)}]{1995JGR...10023443P}
{Priest}, E.~R., \& {D{\'e}moulin}, P. 1995, \jgr, 100, 23443,
  \dodoi{10.1029/95JA02740}

\bibitem[{{Savcheva} {et~al.}(2012){Savcheva}, {Pariat}, {van Ballegooijen},
  {Aulanier}, \& {DeLuca}}]{2012ApJ...750...15S}
{Savcheva}, A., {Pariat}, E., {van Ballegooijen}, A., {Aulanier}, G., \&
  {DeLuca}, E. 2012, \apj, 750, 15, \dodoi{10.1088/0004-637X/750/1/15}

\bibitem[{{Savcheva} \& {van Ballegooijen}(2009)}]{2009ApJ...703.1766S}
{Savcheva}, A., \& {van Ballegooijen}, A. 2009, \apj, 703, 1766,
  \dodoi{10.1088/0004-637X/703/2/1766}

\bibitem[{{Schou} {et~al.}(2012){Schou}, {Scherrer}, {Bush}, {Wachter},
  {Couvidat}, {Rabello-Soares}, {Bogart}, {Hoeksema}, {Liu}, {Duvall}, {Akin},
  {Allard}, {Miles}, {Rairden}, {Shine}, {Tarbell}, {Title}, {Wolfson},
  {Elmore}, {Norton}, \& {Tomczyk}}]{2012SoPh..275..229S}
{Schou}, J., {Scherrer}, P.~H., {Bush}, R.~I., {et~al.} 2012, \solphys, 275,
  229, \dodoi{10.1007/s11207-011-9842-2}

\bibitem[{{Shen}(2021)}]{2021RSPSA.47700217S}
{Shen}, Y. 2021, Proceedings of the Royal Society of London Series A, 477, 217,
  \dodoi{10.1098/rspa.2020.0217}

\bibitem[{{Shen} {et~al.}(2019{\natexlab{a}}){Shen}, {Chen}, {Liu}, {Shibata},
  {Tang}, \& {Liu}}]{2019ApJ...873...22S}
{Shen}, Y., {Chen}, P.~F., {Liu}, Y.~D., {et~al.} 2019{\natexlab{a}}, \apj,
  873, 22, \dodoi{10.3847/1538-4357/ab01dd}

\bibitem[{{Shen} {et~al.}(2018){Shen}, {Liu}, {Song}, \&
  {Tian}}]{2018ApJ...853....1S}
{Shen}, Y., {Liu}, Y., {Song}, T., \& {Tian}, Z. 2018, \apj, 853, 1,
  \dodoi{10.3847/1538-4357/aaa3ff}

\bibitem[{{Shen} {et~al.}(2012{\natexlab{a}}){Shen}, {Liu}, \&
  {Su}}]{2012ApJ...750...12S}
{Shen}, Y., {Liu}, Y., \& {Su}, J. 2012{\natexlab{a}}, \apj, 750, 12,
  \dodoi{10.1088/0004-637X/750/1/12}

\bibitem[{{Shen} {et~al.}(2012{\natexlab{b}}){Shen}, {Liu}, {Su}, \&
  {Deng}}]{2012ApJ...745..164S}
{Shen}, Y., {Liu}, Y., {Su}, J., \& {Deng}, Y. 2012{\natexlab{b}}, \apj, 745,
  164, \dodoi{10.1088/0004-637X/745/2/164}

\bibitem[{{Shen} {et~al.}(2019{\natexlab{b}}){Shen}, {Qu}, {Zhou}, {Duan},
  {Tang}, \& {Yuan}}]{2019ApJ...885L..11S}
{Shen}, Y., {Qu}, Z., {Zhou}, C., {et~al.} 2019{\natexlab{b}}, \apjl, 885, L11,
  \dodoi{10.3847/2041-8213/ab4cf3}

\bibitem[{{Shen} {et~al.}(2022){Shen}, {Zhou}, {Duan}, {Tang}, {Zhou}, \&
  {Tan}}]{2022SoPh..297...20S}
{Shen}, Y., {Zhou}, X., {Duan}, Y., {et~al.} 2022, \solphys, 297, 20,
  \dodoi{10.1007/s11207-022-01953-2}

\bibitem[{{Shen} {et~al.}(2024){Shen}, {Liu}, {Yao}, {Zhou}, {Tang}, {Qu},
  {Zhou}, {Duan}, {Tan}, \& {Ibrahim}}]{2024ApJ...964..125S}
{Shen}, Y., {Liu}, D., {Yao}, S., {et~al.} 2024, \apj, 964, 125,
  \dodoi{10.3847/1538-4357/ad2349}

\bibitem[{{Shen} {et~al.}(2011){Shen}, {Liu}, \& {Liu}}]{2011RAA....11..594S}
{Shen}, Y.-D., {Liu}, Y., \& {Liu}, R. 2011, Research in Astronomy and
  Astrophysics, 11, 594, \dodoi{10.1088/1674-4527/11/5/009}

\bibitem[{{Sturrock}(1966)}]{1966Natur.211..695S}
{Sturrock}, P.~A. 1966, \nat, 211, 695, \dodoi{10.1038/211695a0}

\bibitem[{{Sun} {et~al.}(2013){Sun}, {Hoeksema}, {Liu}, {Aulanier}, {Su},
  {Hannah}, \& {Hock}}]{2013ApJ...778..139S}
{Sun}, X., {Hoeksema}, J.~T., {Liu}, Y., {et~al.} 2013, \apj, 778, 139,
  \dodoi{10.1088/0004-637X/778/2/139}

\bibitem[{{Titov} {et~al.}(2002){Titov}, {Hornig}, \&
  {D{\'e}moulin}}]{2002JGRA..107.1164T}
{Titov}, V.~S., {Hornig}, G., \& {D{\'e}moulin}, P. 2002, Journal of
  Geophysical Research (Space Physics), 107, 1164, \dodoi{10.1029/2001JA000278}

\bibitem[{{T{\"o}r{\"o}k} {et~al.}(2004){T{\"o}r{\"o}k}, {Kliem}, \&
  {Titov}}]{2004A&A...413L..27T}
{T{\"o}r{\"o}k}, T., {Kliem}, B., \& {Titov}, V.~S. 2004, \aap, 413, L27,
  \dodoi{10.1051/0004-6361:20031691}

\bibitem[{{Wang} {et~al.}(2015){Wang}, {Cao}, {Liu}, {Xu}, {Liu}, {Zeng},
  {Chae}, \& {Ji}}]{2015NatCo...6.7008W}
{Wang}, H., {Cao}, W., {Liu}, C., {et~al.} 2015, Nature Communications, 6,
  7008, \dodoi{10.1038/ncomms8008}

\bibitem[{{Wheatland} {et~al.}(2000){Wheatland}, {Sturrock}, \&
  {Roumeliotis}}]{2000ApJ...540.1150W}
{Wheatland}, M.~S., {Sturrock}, P.~A., \& {Roumeliotis}, G. 2000, \apj, 540,
  1150, \dodoi{10.1086/309355}

\bibitem[{{Wyper} {et~al.}(2017){Wyper}, {Antiochos}, \&
  {DeVore}}]{2017Natur.544..452W}
{Wyper}, P.~F., {Antiochos}, S.~K., \& {DeVore}, C.~R. 2017, \nat, 544, 452,
  \dodoi{10.1038/nature22050}

\bibitem[{{Wyper} {et~al.}(2021){Wyper}, {Antiochos}, {DeVore}, {Lynch},
  {Karpen}, \& {Kumar}}]{2021ApJ...909...54W}
{Wyper}, P.~F., {Antiochos}, S.~K., {DeVore}, C.~R., {et~al.} 2021, \apj, 909,
  54, \dodoi{10.3847/1538-4357/abd9ca}

\bibitem[{{Wyper} {et~al.}(2018){Wyper}, {DeVore}, \&
  {Antiochos}}]{2018ApJ...852...98W}
{Wyper}, P.~F., {DeVore}, C.~R., \& {Antiochos}, S.~K. 2018, \apj, 852, 98,
  \dodoi{10.3847/1538-4357/aa9ffc}

\bibitem[{{Zhang} {et~al.}(2012){Zhang}, {Cheng}, \&
  {Ding}}]{2012NatCo...3..747Z}
{Zhang}, J., {Cheng}, X., \& {Ding}, M.-D. 2012, Nature Communications, 3, 747,
  \dodoi{10.1038/ncomms1753}

\bibitem[{{Zhou} {et~al.}(2021{\natexlab{a}}){Zhou}, {Shen}, {Zhou}, {Tang},
  {Duan}, \& {Tan}}]{2021ApJ...923...45Z}
{Zhou}, C., {Shen}, Y., {Zhou}, X., {et~al.} 2021{\natexlab{a}}, \apj, 923, 45,
  \dodoi{10.3847/1538-4357/ac28a0}

\bibitem[{{Zhou} {et~al.}(2021{\natexlab{b}}){Zhou}, {Shen}, {Su}, {Tang},
  {Zhou}, {Duan}, \& {Tan}}]{2021SoPh..296..169Z}
{Zhou}, X., {Shen}, Y., {Su}, J., {et~al.} 2021{\natexlab{b}}, \solphys, 296,
  169, \dodoi{10.1007/s11207-021-01913-2}

\end{thebibliography}

\end{document}